# The intensity and evolution of the extreme storms in January 1938

Hisashi Hayakawa (1 – 4)*, Kentaro Hattori (5), Alexei A. Pevtsov (6 – 7), Yusuke Ebihara (8), Margaret A. Shea (9), Ken G. McCracken (10), Ioannis A. Daglis (11 – 12), Ankush Bhaskar (13), Paulo Ribeiro (14), Delores J. Knipp (15 – 16)

(1) Institute for Space-Earth Environmental Research, Nagoya University, Nagoya, 4648601, Japan
(2) Institute for Advanced Researches, Nagoya University, Nagoya, 4648601, Japan
(3) UK Solar System Data Centre, Space Physics and Operations Division, RAL Space, Science and Technology Facilities Council, Rutherford Appleton Laboratory, Harwell Oxford, Didcot, Oxfordshire, OX11 0QX, UK
(4) Nishina Centre, Riken, Wako, 3510198, Japan
(5) Graduate School of Science, Kyoto University, Kyoto, 6068501, Japan
(6) National Solar Observatory, 3665 Discovery Drive, 3rd Floor, Boulder, CO 80303, USA
(7) Central Astronomical observatory of Russian Academy of Sciences at Pulkovo, Saint Petersburg, 196140, Russia
(8) Research Institute for Sustainable Humanosphere, Kyoto University, Uji, 6110011, Japan
(9) SSSRC, 100 Tennyson Avenue, Nashua, NH 03062, USA
(10) 64 Burradoo Rd., Burradoo, 2576, NSW, Australia
(11) Department of Physics, National and Kapodistrian University of Athens, Athens, 15784, Greece
(12) Hellenic Space Center, Athens, 15231, Greece
(13) NASA Goddard Space Flight Center, Greenbelt, MD, United States
(14) University of Coimbra, CITEUC, Geophysical and Astronomical Observatory, Coimbra, 3040-004, Portugal
(15) Smead Aerospace Engineering Sciences Department, University of Colorado Boulder, Boulder, CO 80309, USA
(16) High Altitude Observatory, National Center for Atmospheric Research, Boulder, CO 80307, USA

* hisashi@nagoya-u.jp

**Abstract**

Major solar eruptions occasionally direct interplanetary coronal mass ejections (ICMEs) to Earth and cause significant geomagnetic storms and low-latitude aurorae. While single extreme storms are of significant threats to the modern civilization, storms occasionally appear in sequence and, acting synergistically, cause 'perfect storms' at Earth. The stormy interval in January 1938 was one of such cases. Here, we analyze the contemporary records to reveal its time series on their source active regions, solar eruptions, ICMEs, geomagnetic storms, low-latitude aurorae, and cosmic-ray (CR) variations. Geomagnetic records show that three storms occurred successively on 17/18 January (Dcx ≈ −171 nT) on 21/22 January (Dcx ≈ −328 nT) and on 25/26 January (Dcx ≈ −336 nT). The amplitudes of the cosmic-ray variations and sudden storm commencements show the impact of the first ICME as the largest (≈ 6% decrease in CR and 72 nT in SSC) and the ICME associated with the storms that followed as more moderate (≈ 3% decrease in CR and 63 nT in SSC; ≈ 2% decrease in CR and 63 nT in SSC). Interestingly, a significant solar proton event occurred on 16/17 January and the Cheltenham ionization chamber showed a possible ground level enhancement. During the first storm, aurorae were less visible at mid-latitudes, whereas during the second and third storms, the equatorward boundaries of the auroral oval were extended down to 40.3° and 40.0° in invariant latitude. This contrast shows that the initial ICME was probably faster, with a higher total magnitude but a smaller southward component.

**1. Introduction**

Large and complex sunspot groups occasionally trigger solar flares and launch sequential coronal mass ejections (CMEs) into space, where they are identified as interplanetary coronal mass ejections (ICMEs) (Gopalswamy *et al.*, 2005; Tsurutani *et al.*, 2014; Liu *et al.*, 2019). ICMEs with a southward magnetic field component that impact Earth typically initiate a significant geospace magnetic storm with an equatorward expansion of the auroral oval (Gonzalez *et al.*, 1994; Yokoyama *et al.*, 1998; Daglis *et al.*, 1999; Gopalswamy *et al.*, 2007).

The intensity of geomagnetic storms has been measured through the Dst index, as a proxy of the terrestrial ring-current intensity, on the basis of variability of the horizontal intensity of four mid-latitude stations with geomagnetic latitudinal weighting: Kakioka, Hermanus, San Juan, and Honolulu (Sugiura, 1964; Gonzalez *et al.*, 1994; Daglis *et al.*, 1999; Daglis, 2006; WDC for Geomagnetism at Kyoto *et al.*, 2015). Since the beginning of the Dst index measurement in 1957, the largest geomagnetic storm was recorded in March 1989 (with a record value of most negative Dst = −589), during which significant low-latitude aurorae and serious blackouts were recorded (Allen *et al*., 1989; Boteler, 2019). Extending our investigations back to the beginning of systematic magnetic measurements in the mid 19th century, we note other intense geomagnetic storms such as those in September 1859, February 1872, and May 1921 (Tsurutani *et al*., 2003; Silverman, 2006; Cliver and Dietrich, 2013; Hayakawa *et al*., 2018, 2019; Hapgood, 2019; Love *et al*., 2019).

Understanding such intense storms is more than just an academic concern, as the occurrence of such storms represents a significant risk to our modern civilization, because of our increasing dependency on technology infrastructure that is vulnerable to various aspects of space storms. Among them, the 1989 March storm and other major storms have seriously affected human civilizations with their resultant geomagnetically induced currents (*e.g*., Allen *et al*., 1989; Boteler, 2019). Modern consequences of the superstorms in September 1859 and May 1921 have been considered even more catastrophic and have been studied intensively (Daglis, 2001; Daglis, 2005; Baker *et al*., 2008; Hapgood, 2018; Riley *et al*., 2018).

Such ICMEs may become even more geo-effective when a series of them are launched from a single source sunspot active region. For such a sequence of events, initial ICMEs sweep the interplanetary space allowing the following ICMEs to decelerate less (Tsurutani and Lakhina, 2014; Tsurutani *et al*., 2014; Shiota and Kataoka, 2016). This was the case with the Halloween sequence in 2003 October (Gopalswamy *et al*., 2005; Mannucci *et al*., 2005; Shiota and Kataoka, 2016). Close inspections of the time series of the extreme storms and superstorms show that they occasionally consist of multiple storms within several days (*e.g.*, Silverman, 2006; Cid *et al*., 2014; Knipp *et al.*, 2016; Lefèvre *et al*., 2016; Boteler, 2019; Hayakawa *et al*., 2017, 2019; Hattori *et al*., 2019; Love *et al*., 2019).

The geomagnetic storm on 25 January 1938 was such a case. Its intensity was ranked 10th in the observations of Greenwich-Abinger magnetograms (Jones, 1955, p. 79) and in the 33rd in the *aa* index in 1868 – 2010 (Lefevre *et al*., 2016), and accompanied with splendid auroral display throughout Europe even down to Gibraltar, Sicily, and Greece (Störmer, 1938; Anon., 1938; Carpiperis, 1956; Correia and Ribeiro, 1996), and a series of global decreases in the cosmic-ray intensity (Forbush Decreases) detected by ionization chambers (Forbush, 1938; Hess *et al*., 1938). The 25 January storm was the third in a sequence of events that produced four sudden impulses, three of which created storm sudden commencements and associated geomagnetic storms in a nine-day interval. Among them, two storms on 21/22 January with less intensity and 25/26 January with a higher intensity, have been highlighted as twin occurrences with great auroral displays and have been compared to the superstorms around the Carrington event in August and September 1859 (Silverman, 2006; see also Hayakawa *et al.*, 2019).

Since these geomagnetic storms occurred long before the development of the Dst index in 1957, these events and the time series of the associated solar and terrestrial phenomena have been in want of quantitative evaluations and detailed analyses to be compared with the major storms during the modern instrumental observations. Therefore, in this article, we have aimed at reconstructing their time series from onset at the source active region, characterizing the source flares and ICMEs, to the intensity and time series of the geomagnetic storms, the low-latitude aurorae, and the solar cosmic-ray variability, on the basis of the contemporary observational records.

## 2. Solar Eruptions

Solar Cycle 17 reached its maximum in April 1937 (Table 1 of Hathaway, 2015; Figure 2 of Clette and Lefèvre, 2016), culminating in a giant sunspot group crossing the solar central meridian on April 24. The sunspot activity has been regularly monitored at the Mt. Wilson Observatory (*e.g.*, Pevtsov *et al.*, 2019) and Tashkent Observatory (Slonim and Ushakova, 1938; Slonim, 1939) around this period and recorded as daily sunspot magnetic field measurements and daily sunspot drawings. On their basis, over the following 8 – 9 months, the sunspot activity went through multiple episodes of enhanced-diminished activity with several giant sunspots visible with an unaided eye forming mostly in the northern solar hemisphere. Thus, for example, the sunspot activity had weakened in May 1937, but by the mid-June, several active regions begun developing in both solar hemispheres. On 29 July 1937, a sunspot group with two giant naked-eye sunspots situated in the northern solar hemisphere crossed the central meridian. Over following months, the activity had weakened again. Another giant sunspot group had developed in the Northern hemisphere and crossed the central meridian around 4 October 1937, and then again, November-December was relatively calm in its sunspot activity. Disk-passage of the giant spots, which developed during this period of 8 months was accompanied by major geomagnetic storms. In 1937, the sunspot activity exhibited strong hemispheric asymmetry with the Northern hemisphere being the most active. Interestingly, the majority of giant sunspot groups, which were observed on the Sun in 1937, had developed in the range of antipodal Carrington longitudes of 175°–195° and 350°–355° (Moisejev, 1939a), which may be associated with solar active longitude (*e.g.*, Becker, 1955; Bumba and Howard, 1965; Haurwitz, 1968; Berdyugina and Usoskin, 2003; Sudol and Harvey, 2005; Usoskin *et al.*, 2007). A giant sunspot active region (AR)

RGO[1] 12673 (MWO[2] 5726) appeared on the solar disk at the East limb in the Northern hemisphere on 12 January. It crossed the central meridian on 18 January, and disappeared behind the West limb on 24 January (Figure 1). This was the seventh largest sunspot region by mean area between 1874 – 1954 (Jones, 1955) and was a recurrent region, existing over three consecutive solar rotations. The Carrington longitudinal position of this region was close to one of the "active longitudes" that produced a number of giant sunspots in 1937. The AR grew to 3627 millionths of the solar hemisphere (msh)[3]. The main sunspot of this group was a shapeless conglomerate of multiple umbrae with the opposite polarity magnetic fields surrounded by a common penumbra and had area of about 3361 msh (Moisejev, 1939a; Kurochkin, 1939; Slonim and Ushakova, 1938; Slonim, 1939; see the daily sunspot magnetic field observations and daily sunspot drawings at the Mount Wilson Observatory (*e.g.*, Pevtsov *et al.*, 2019)).

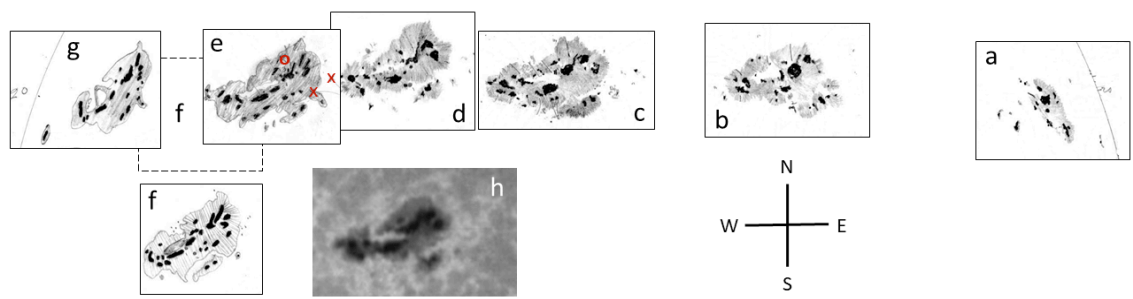

Figure 1: Evolution of active region MWO5726 during its disk passage in January 1938: (a) 13 January, 17:50 UT, (b) 16 January, 16:20 UT, (c) 18 January, 16:15 UT, (d) 19 January, 19:00 UT, (e) 20 January, 11:30 UT, (f) 21 January, 19:45 AM, (g) 22 January, 19:10 UT, and (h) spectroheliogram in Ca K1 for the location corresponding to panel (d) on 19 January. The locations of panels (a-g) correspond to their heliographic coordinates on the solar disk for the day of observation. The diameter of solar disk can be inferred from the position of solar limbs shown in panels (a) and (g). Panel (f) is shifted from its true location (shown as dashed square). Red crosses in panel (e) mark the location of two bright Hα kernels, and red circle is the location of large Doppler velocity measured in Hα, both measured at 00:00 UT (21 January). The measured Doppler shift (2 Å) corresponds to the Doppler velocity of $\approx 90$ km/s.

---

[1] Royal Greenwich Observatory
[2] Mt. Wilson Observatory
[3] https://solarscience.msfc.nasa.gov/greenwch.shtml

During its disk passage, this AR produced a number of solar flares including four major flares with their Hα importance = 3 as well as numerous weaker flares of importance 1 and 2, as recorded in *Quarterly Bulletin of Solar Activity* (hereafter QBSA) (D'Azambuja, 1938, p. 124). Note that this importance indicates the Hα flaring area (1 = 100 – 250 msh; 2 = 250 – 600 msh, and 3 = 600 – 1200 msh; see Švestka, 1976, p. 14), whereas the relative brightness has been annotated with B for bright, N for normal, and F for faint. Optical flares in Hα are usually accompanied by radio and X-ray bursts, and occasionally by high-energy particle emissions.

So far we have at least 4 major flares (= class 3) in this interval. The first major flare was observed on 14 January at around 4:40 – 5:30 UT with a spectrohelioscope at Watheroo, when this AR was positioned at N10E45 (south of the main sunspot). The second major flare was observed on 20 January at around 18:20 – 21:27 UT at Mt. Wilson, when this AR was positioned at N18W30. The maximum of Hα flare was recorded at ≈ 19:52 UT. The third and fourth major flares[4] were observed on 24 January 1938: one during 3:00 – 3:40 UT according Watheroo observations and the other at around 5:12 – 7:00 UT based on spectrohelioscope observations from Canberra from the AR at N22W85 to N22W80. The flare on 24 January is especially noted as "in connection with an eruptive protuberance" (D'Azambuja, 1938, p. 124) and listed as one of great flares that occurred in an active longitude complex (Haurwitz, 1968). As their flare importance were classified as 3, it is considered that the flare area in Hα reached 600 – 1200 msh (*e.g.*, Švestka 1976, p. 14). Moreover, it is quite possible that not all the flare activities from this region were recorded in the QBSA. Thus, for example, MWO drawing taken on 21 January (Figure 1, panel (e)) contains notes about two bright Hα kernels, as well as a measurement of a significant shift in Hα spectral line. The shift corresponds to a significant (≈ 90 km/s) Doppler velocity consistent with an eruptive event. The time of these measurements (00UT on 21 January) is about 4 hours after the last eruptive event on 20 January and about 2 hours before the first eruptive event on 21 January listed in the QBSA. Thus, it must be a different flare of unknown importance.

Given the strength (importance) of optical (Hα) flares in AR MWO 5726, there is a high probability that this region produced major X-ray flares. For example, Hayakawa *et al*. (2020a) used NOAA lists of flares to demonstrate that in 96% cases, the H-alpha flare of importance 3 is accompanied with either X-class (66%) or M-class (30%) flares. Major X-ray (X- and M-class) flares are almost always accompanied with interplanetary coronal mass ejections (ICMEs), as shown from the statistical studies with the LASCO (Large Angle and Spectrometric Coronagraph; Brueckner *et al.*, 1995) observations on the SOHO (Solar and Heliospheric Observatory; Domingo *et al.*, 1995) mission during 1996 – 2010 (Youssef, 2012; 90% of all X-class flares and 30% of all M-class flares are accompanied by ICMEs). Indeed, shortly after these flares, three storm sudden commencements (SSCs) with significant amplitude were recorded on 16 January (22:36 UT, 72 nT), 22 January (02:42 UT, 63 nT), and 25 January (11:51 UT, 63 nT) at Kakioka Observatory[5]. Their time lags show that these ICMEs responsible for these SSCs reached the Earth 65.9 h, 30.3 h, and 32.9 h, respectively, after the occurrence of the Hα flare. These time lags yield average velocities as 630 km/s, 1370 km/s, and 1260 km/s, respectively.

---

[4] Kurochkin (1939) mentions an Hα eruption in this active region on 23 January at 7:20 – 7:30 UT although without a reference to this flare importance.

Several active regions were present on the solar disk during the disk passage of active region MWO 5726, and some of these regions also produced Hα flares albeit none of the other active regions produced flares of importance 3 as recorded in the QBSA. For example, on January 14, a relatively large Hα flare of importance 2+ was observed at N25W35, close to AR MWO 5719. The CMEs originating West of central meridian are more likely to be geoeffective as compared with the CMEs originating east of Sun's central meridian (Gopalswamy *et al.*, 2007). Thus, the geomagnetic storm on January 17 may be related to a CME originating from region MWO 5719, not from region MWO 5726. Nevertheless, both eruptions occurred at about the same time, and thus, this difference in the source region is not important for estimating the time between the CME liftoff and the onset of the geomagnetic storm.

However, the reported flares were probably no more than a part of the entire solar flare activity in this interval. Notably, we have 10 radio fadeouts reported in the Eastern USA (Table III of Gilliland, 1938) and several SFEs in this interval (*e.g.*, Bartels, 1939; Yokouchi, 1953). Therefore, for the first geomagnetic storm on 17 January, we have two more possible scenarios, as relatively large solar flare effects (SFEs: see *e.g.*, Curto (2020) for a review) were reported at Cheltenham, Tucson, San Juan, and Huancayo at ≈ 17:07 UT on 15 January 1938 and at Watheroo, Honolulu, and Kakioka at ≈ 0:40 UT on 16 January 1938 (see *e.g.*, Figure 2; Bartels *et al.*, 1939; Yokouchi, 1953; Cliver and Svalgaard, 2004). The latter was especially accompanied with a notable SFE at Apia (56 nT; Wadsworth, 1938), which was probably enhanced with equatorial electrojet due to its proximity to the geomagnetic equator (see *e.g.*, Rastogi *et al.*, 1997). These ionospheric disturbances indicate an intense X-ray flare overlooked in the flare patrols at that time (*c.f.*, D'Azambuja, 1938, p. 124). In these cases, as the ICME transit times were ≈ 29.5 h and ≈ 21.9 h (for the latter, see also Bartels *et al.*, 1939; Cliver and Svalgaard, 2004), their average ICME velocities are estimated ≈ 1400 km/s and ≈ 1900 km/s, respectively. These scenarios will be further analysed on the basis of the cosmic-ray variability recorded in the contemporary ionization chambers in Section 5.

---

[5] http://www.kakioka-jma.go.jp/obsdata/Geomagnetic_Events/Events_index.php

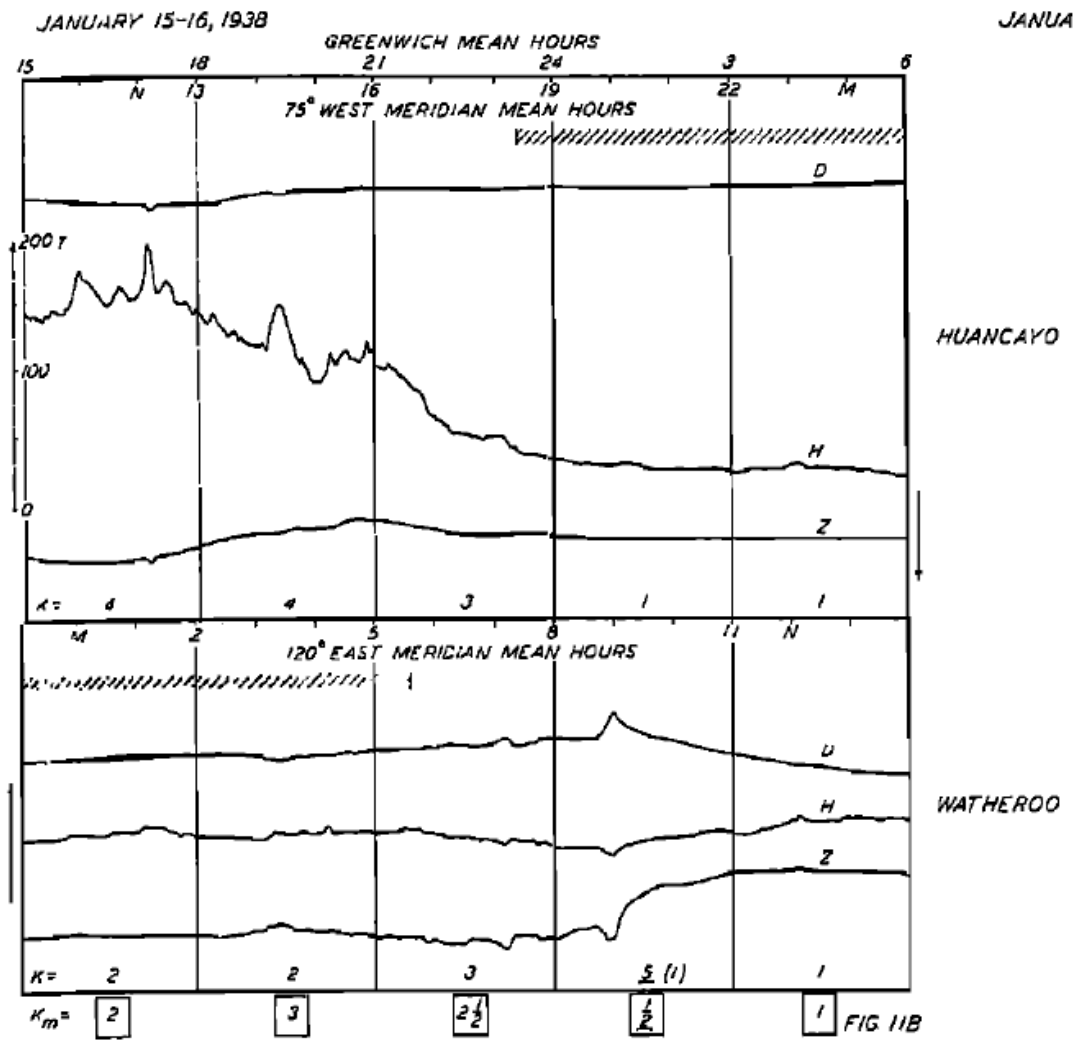

Figure 2: Traces with the SFEs at 17:07 UT on 15 January recorded at Huancayo Observatory (upper panel) and at 00:40 UT on 16 January recorded at Watheroo Observatory (lower panel), adopted from Bartels *et al.* (1939).

**3. Geomagnetic Storms**

Upon arrival, these ICMEs caused three consecutive intense geomagnetic storms. Usually, the Dst (disturbance storm time) index is used to characterize the storm intensity and time series to follow the development of the ring current. However, the official Dst index became available only in 1957 and hence does not cover these storms. Instead, we have used the Dcx index, a so-called corrected and extended version of the Dst index created by the space climate team at the University of Oulu. The Dcx index starts in 1932 (Karinen and Mursula, 2005, 2006; Mursula *et al.*, 2008). As Hermanus Observatory started operation only after 1941, the Dcx index in 1938 has been reconstructed with the observations from the neighboring magnetic station of Cape Town as replacement of Hermanus data. We also used the hourly data for these four stations (Kakioka, Cape Town, San Juan, and Honolulu), and confirmed that they are free from scale-off issues, and cross-checked the calculated results[6].

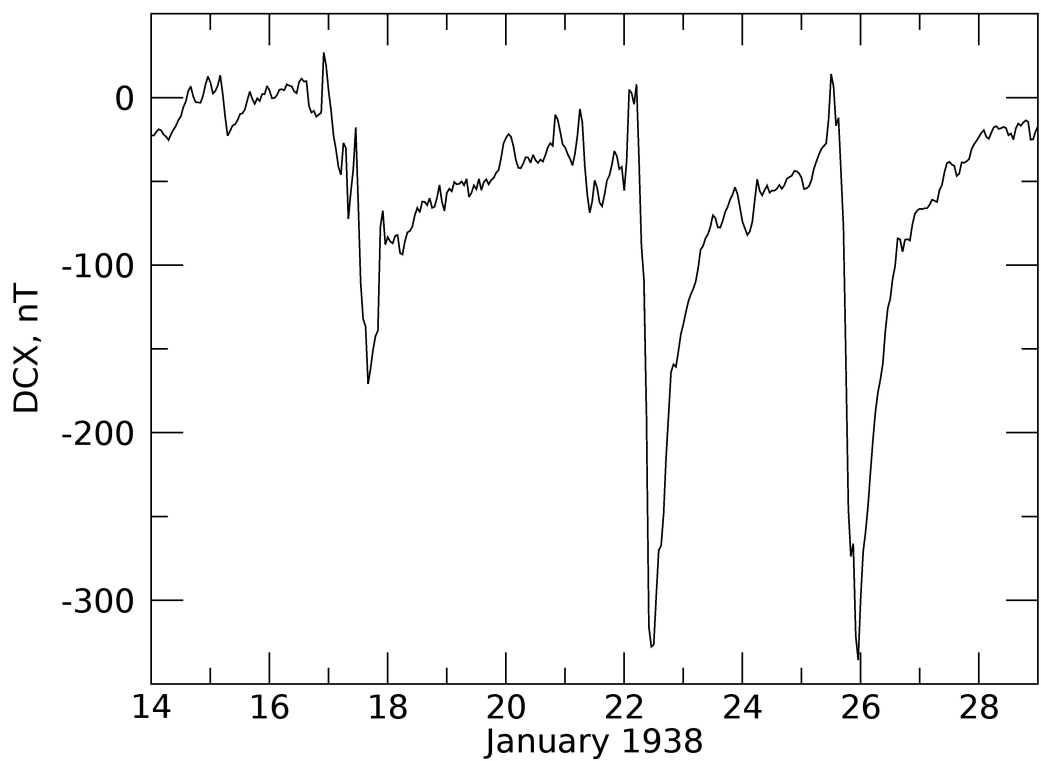

Figure 3: The hourly Dcx index during the period 14 – 29 January 1938.

---

[6] http://wdc.kugi.kyoto-u.ac.jp/caplot/index.html

Figure 3 shows the time series of the hourly Dcx index during 14 – 29 January 1938. Three major storms were recorded in this interval as identified from the negative excursions with large amplitudes. The first storm peaked at 16 UT on 17 January 1938 with its maximum negative Dcx ≈ −171 nT after its SSC at 22:36 UT on 16 January. The second and third storms are often considered as twin storms (see Jones, 1955; Lefèvre *et al*., 2016) and peaked at 11 UT on 22 January with Dcx ≈ −328 nT and 23 UT on 25 January with Dcx ≈ −336 nT.

Jones (1955) identifies the geomagnetic storm on 25 January 1938 as among the top eleven storms of the 112 great storms listed between 1874-1954. Of particular note was the apparent "mis-identification" of the solar source in the Greenwich Catalog storm tabulation (Jones, 1955, pp. 77-81) because of the statistical limits of the distance of Group 12673 from the solar central meridian.

The storm intensity of the second and third storms are comparable to the major magnetic storm on 12 November 1960 (Dst = −339 nT) and only slightly weaker than the Halloween storms on 29/30 October 2003 (Dst = −354 nT), and on 30/31 October 2003 (Dst = −383 nT), as well as the extreme storm on 25/26 May 1967 (WDC for Geomagnetism at Kyoto *et al*., 2015; see also Gopalswamy *et al*., 2005; Knipp et al. 2016; Lockwood *et al*., 2019). This amplitude is certainly in an extreme category (Meng *et al*., 2019), while it does not go beyond the threshold of Dst (or Dst*[7] before 1956) being equal to −500 nT (Cliver and Dietrich, 2013; Hayakawa *et al*., 2019, 2020a, 2020b). Caveats must be noted for its uncertainty, as the Dst index and Dcx index are slightly different in the calculation procedure and amplitudes of specific storms in these indices vary up to ≈ 44 nT (see Karinen and Mursula, 2006; Mursula *et al*., 2008; Riley, 2017).

**4. Low- and Mid-Latitude Aurorae**

The auroral oval expands equatorward during major geomagnetic storms (Kamide and Winningham, 1977; Yokoyama *et al*., 1998; Shiokawa *et al*., 2005). It was also the case during this stormy interval, especially during the storms on 21/22 and 25/26 January (Silverman, 2006). In particular the last storm on 25/26 January was characterized by aurorae "seen over practically the whole of Europe, and as far south as Gibraltar and Sicily" (Anon. 1938, p. 232).

---

[7] Here, we denote equivalent Dst estimates with alternative stations as Dst*.

We have extended investigations on the visual auroral reports across the United States, the USSR, Australia, and New Zealand, using the summary of auroral observations for years 1938-1939 published by Kurochkin (1939), the *Meteorological Data in the United States* (U. S. Department of Agriculture Weather Bureau, 1938), and local reports and newspapers. These reports show that the aurorae extended equatorward for the three storms on 17/18, 21/22, and 25/26 January 1938 (Figure 3). Reports in the United States show the significant latitudinal expansion of auroral visibility during these storms (Figure 4). The *Meteorological Data in the United States* reported the auroral visibility during these three geomagnetic storms. Even on 17/18 January, the aurorae were visible down to Cheyenne in Wyoming (N41°08′, W104°49′; 50.0° MLAT). The aurorae were predominantly observed in the central United States on 21/22 January and in the eastern United States on 25/26 January.

This was also the case with Australia and New Zealand. The local newspapers show significant auroral visibility on 17/18, 22/23, and 25/26 January (Figure 4). In these countries, the aurorae were most splendid on 22/23 January and visible down to Norfolk Island (S29°02′, E167°57′; −34.9° MLAT). Despite the reduced significance, aurorae were reported during the other nights: down to Wellington (S41°17′, E174°46′; −45.6° MLAT) on 17/18 January; and down to Manilla (S30°45′, E150°43′; −39.4° MLAT) on 25/26 January.

In the USSR, during the month of January, there were in total 24 auroral nights reported in the USSR. Five of those auroral displays were so bright that they produced multiple reports from the (geographic) mid-latitudes. All auroras were concurrent with the major geomagnetic storms and were accompanied by interruptions in radio and telegraph communications. Kurochkin (1939) provided a summary of the aurora observations based on the numerous eye-witness letters received by the local newspapers, the Moscow Planetarium, the Division of the Astronomical-Geodetical Society of the USSR, and the Sternberg Astronomical Institute. We supplemented this summary by the additional reports from regional newspapers. The approximate geographic locations of the aurora reports are shown in Figure 4.

Aurora on 17 January 1938 was observed between 13 UT and 22 UT as a diffuse glow near the horizon with pillars growing out of the glow. The pillars were moving from East to West. The auroral colors in the eastern regions of the USSR were whitish at first, but turned into the reddish hues later. In western regions, the aurora was mostly red in hue. On 22 January, the aurora was first observed in the USSR Far East at the latitude of about 48°. The aurora had a shape of diffuse clouds with the rays of greenish and purple color. The aurora observed in the western part of country was also in the shape of diffuse clouds, but without rays. The auroral features appeared moving from East to West. The strongest auroral activity was observed during 25 – 27 January, and it had been seen as far South as the Crimean Peninsula (N45°). On 25 January, the aurora was reported as a series of white bands, which later turned into bright-red diffuse clouds. On 26 January, in the eastern part of the country, the reports of aurora described it as the series of pillars of red color. In the western regions, the aurora has been seeing as a red glow low in the horizon. The brightness of aurora on 27 January was weaker as compared with the previous nights. At the beginning the colors were reported as light-red, but gradually changed to a glow of a whitish color. The auroral patterns were moving from West to East (Kurochkin, 1939). The description of the observed patterns and their colors is in a good agreement with the examples of aurora drawings shown in Figure 5.

Figure 4 shows the geographic distributions of visual auroral reports on 17/18, 21/22 and 25/26 January 1938. As shown here, the auroral visibility actually extended much more equatorward than observed in the European sector. In comparison with Silverman (2006), our investigations show further observations in the northern Japan including Southern Sakhalin Island (currently under rule in the Russian Federation) as well as a large longitudinal extension of aurora observations across the former USSR during both of these storms (see *e.g.*, Figure 4), Northern Australia and North-Eastern China (mainly Manchuria at that time) on 21/22 January, and Greece and French Northern Africa (modern-time Tunisia, Algeria, and Morocco) on 25/26 January. On the basis of the magnetic field model IGRF12 (Thébault *et al*., 2015), we have calculated the geomagnetic pole, which was located at N78°29′, W068°27′ in 1938. Magnetic latitudes (MLATs) of given observational sites were calculated with their angular distance from the geomagnetic pole. Accordingly, the most equatorial auroral visibility was confirmed as 29.5° MLAT at Morioka in Japan (N39°42′, E141°09′; Morioka Observatory, 1954, p. 117) on 21/22 January and as −29.9° at the vessel Nadrana off South Africa (S31°50′, E015°30′; Anon., 1939). As the geomagnetic dipole strength is continuously declining (*e.g.*, Korte and Constable, 2011), these auroral expansions may have required a slightly stronger solar wind driver than expected from the modern statics (Siscoe and Christopher, 1975; Ebihara and Tanaka, 2020).

17/18 Jan. 1938

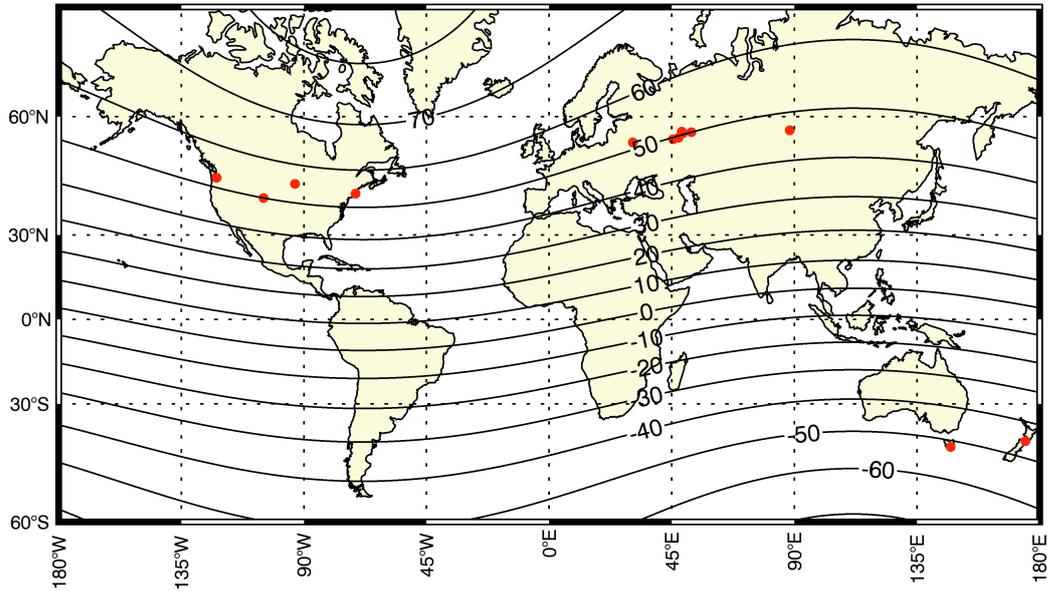

21/22 Jan. 1938

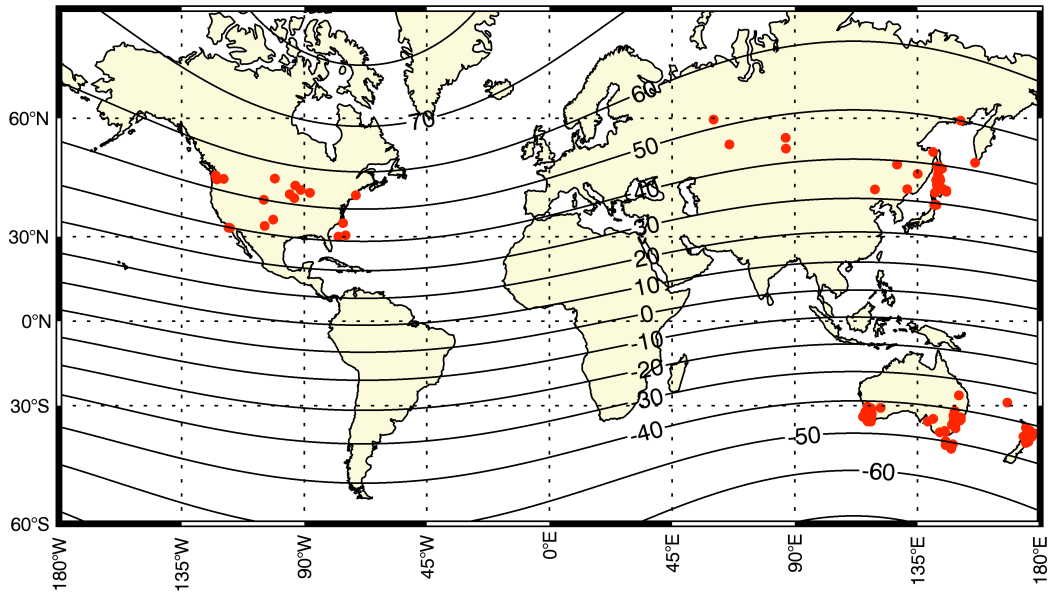

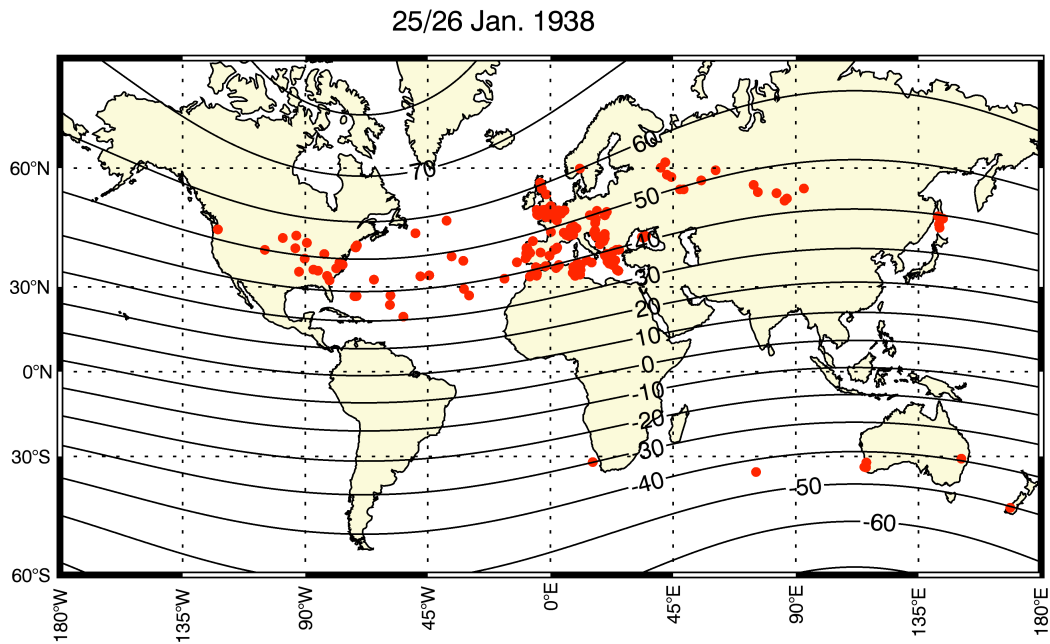

Figure 4: Reported auroral visibility on 17/18 (above), 21/22 (middle), and 25/26 (below) January 1938[8]. The reports taken at |λ| < 40° MLAT have been primarily investigated here and added to the data points in Silverman (2006), U. S. Department of Agriculture Weather Bureau (1938), the USSR accounts (*e.g.*, Kurochkin, 1939), and newspapers in Australia and New Zealand. The observational sites are shown with the red dots. The contour lines indicate MLATs with interval of 10° on the basis of the IGRF 12 model (Thébault *et al.*, 2015).

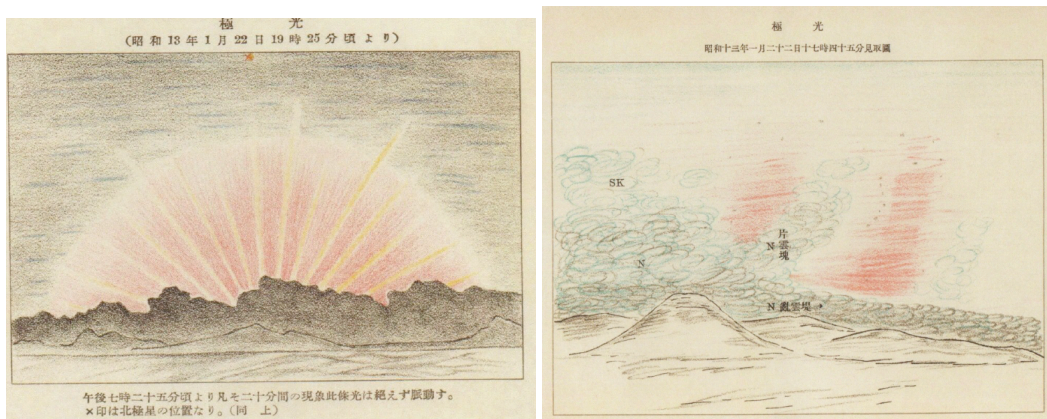

Figure 5: Auroral drawings on 22 January 1938: at Keramui (33.8° MLAT; N43°40′, E145°33′) at 19:25 LT (left; Kisho Yoran, 1938a) and Paramushir (41.9° MLAT; N50°40′, E156°07′) at 19:45 LT (right; Kisho Yoran, 1938b). The Keramui drawing shows red background and yellow streaks and

---

[8] https://www.kwasan.kyoto-u.ac.jp/~hayakawa/data/1938/1938_Jan_aurora_data.txt

mention perpetual pulsation of the auroral emissions. The Paramushir drawing shows vertical auroral extension. The reddish component and yellow streaks are considered as Oxygen emissions at 630.0 nm with some mixture of Oxygen emissions at 557.7 nm.

Among these storms, the aurorae on 21/22 and 25/26 January were especially notable in terms of their equatorial extent and were visible even below 40° MLAT. The MLATs of the auroral visibility ($|\lambda| < 40°$ MLAT) are shown in Figure 6 together with the time series of the Dcx index. This figure shows that the aurora was visible at low-latitudes ($|\lambda| < 40°$ MLAT) near the minima of the Dcx index (Dcx ≤ −200 nT), namely in the late main phase to the beginning of the recovery phase.

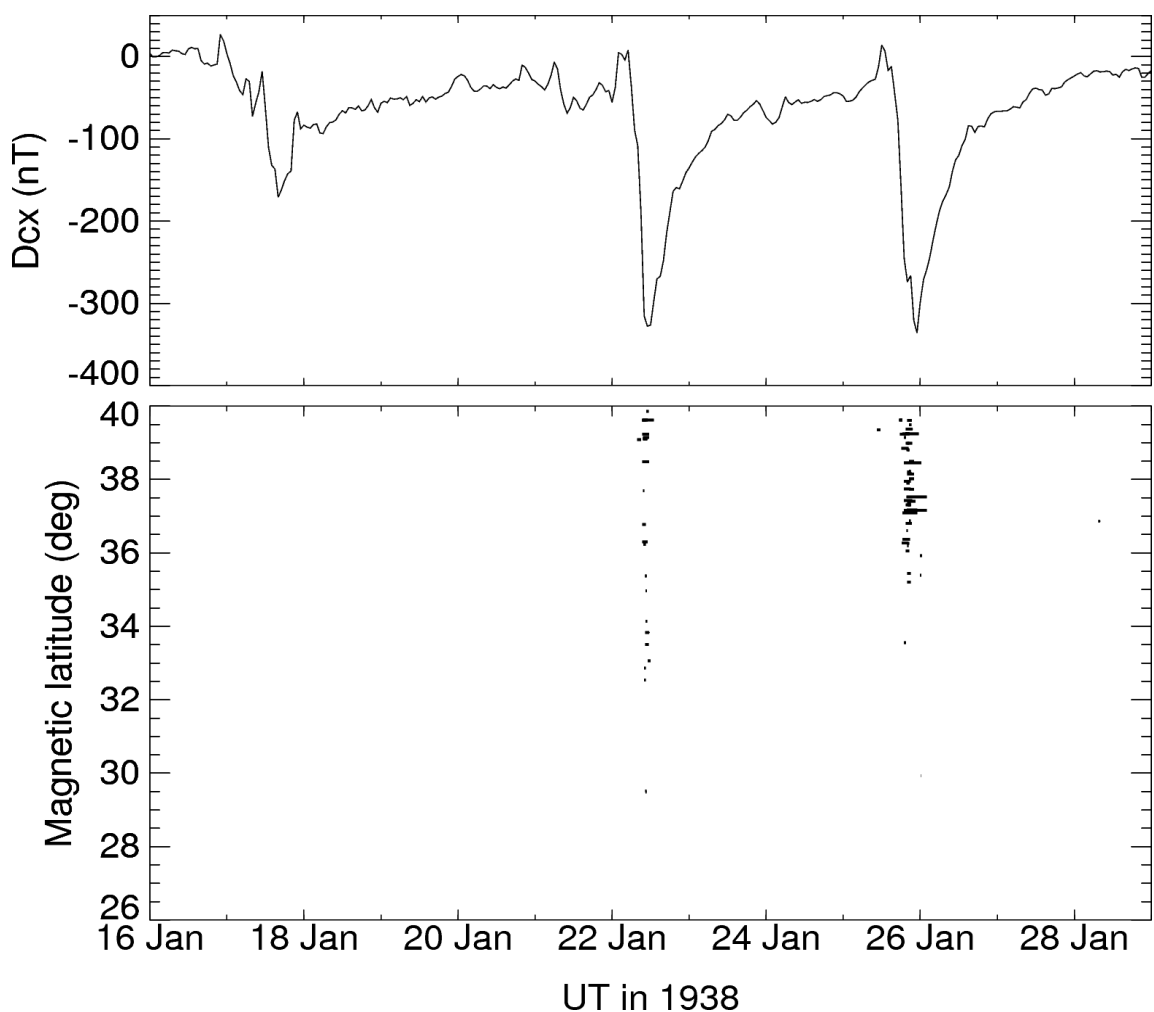

Figure 6: The hourly Dcx index (nT) enlarged from Figure 3 and the low-latitude auroral visibility as a function of $|\lambda|$ and time[9].

---

[9] https://www.kwasan.kyoto-u.ac.jp/~hayakawa/data/1938/1938_Jan_aurora_data.txt

Reports with auroral elevation angle enable us to estimate the equatorial boundary of the auroral oval in combination with their MLATs. Here, we compute the invariant latitude (ILAT) of the footprint of its magnetic field line along which the auroral electrons precipitate (O'Brien *et al.*, 1962; McIlwain, 1966), assuming the auroral elevation as ≈ 400 km (Roach *et al.*, 1960; Ebihara *et al.*, 2017). During the storm on 21/22 January, the aurorae were reported overhead at Noto in Sakhalin Island (39.2° MLAT; N49°16′, E144°00′; Karafuto Department Observatory, 1939, p. 6; see Figure 5). Accordingly, its equatorial boundary of the auroral oval is computed as 40.3° ILAT. These results agree with one another and with the report of most equatorial visibility on 21/22 January as well. In this case, the aurorae were probably visible up to ≈ 20° in elevation angle at Morioka, at the most equatorial observational site (29.5° MLAT; N39°42′, E141°09′; Morioka Observatory, 1954, p. 117). On the other hand, during the 25/26 January storm, the aurorae were reported up to 45° in elevation angle at Patras in Greece (37.3° MLAT; N38°15′, E021°44′; Η ΔΡΑΣΙΣ (I Drasis, 1938), 1938-1-28, p.1). On this basis, its equatorial boundary of the auroral oval has been computed as 40.0° ILAT and compares well with that on 25/26 January.

**5. Cosmic-Ray Variations**

Cosmic ray measurements at Earth in 1938 were primarily those from ionization chambers or an occasional balloon flight (Shea and Smart, 2000), whereas their calibrations with the neutron-monitor data are still challenging (McCracken, 2007; Usoskin *et al.*, 2011; Shea and Smart, 2019). Ionization chambers at Cheltenham and Cambridge in the USA, Christchurch in New Zealand, Teoloyucan near Mexico City in Mexico, Huancayo in Peru, and Hafeakala in Austria continuously monitored the cosmic ray intensity. In addition to those detectors, the cosmic ray group in the Netherlands had an ionization chamber onboard a ship with measurements being made at different locations dependent on the route traveled (Shea and Smart, 2000).

The availability of cosmic ray particles at the top of the atmosphere is dependent on the cutoff rigidity of the detector location. The polar regions are accessible by MeV protons whereas it takes a high energy particle (≈ 15 GeV) to penetrate the earth's magnetic shield to the equatorial atmospheres. The shielded ionizations chambers respond primarily to secondary muons generated by the incident particles creating a nuclear cascade in the atmosphere. In order for this short-lived muon[10] to survive its transit through the atmosphere to sea level, it must be the product of an incoming nuclei > 4 GeV (equivalent to a rigidity of ≈ 4.85 GV). This relativistic energy may be slightly lower for detectors at altitude as the transit path through the atmosphere is slightly shorter.

The detectors at Cambridge, Cheltenham, Christchurch and Hafelekar, with geomagnetic cutoff rigidity values < 4.85 GV would detect only the muons created by the cosmic radiation above ≈ 4.85 GV. The other two detectors at Teoloyucan and Huancayo would detect only the muons created by cosmic rays above their respective cutoff rigidity (See Table 1).

Table 1: Locations of the cosmic-ray ionization chambers in January 1938, derived from Table I of Shea and Smart (2000).

| Name* | Latitude | Longitude | Altitude (m) | Cutoff Rigidity (GV)** |
|---|---|---|---|---|
| Cambridge, USA (Boston) | N42°22′ | W071°05′ | 3 | 1.47 |
| Cheltenham, USA | N38°42′ | W076°48′ | 72 | 2.04 |
| Christchurch, NZ | S43°30′ | E172°36′ | 8 | 2.82 |
| Hafelekar, Austria | N47°19′ | E011°23′ | 2290 | 4.26 |
| Teoloyucan, Mexico | N19°12′ | W099°12′ | 2285 | 10.09 |
| Huancayo, Peru | S12°18′ | W075°20′ | 3350 | 13.67 |

* Some researchers refer to these locations by the alternate name given in parenthesis.

** Vertical cutoff rigidity using the IGRF geomagnetic field coefficients appropriate for 1940.

---

[10] The half life of a muon in the laboratory frame is 2.2 micro-seconds.

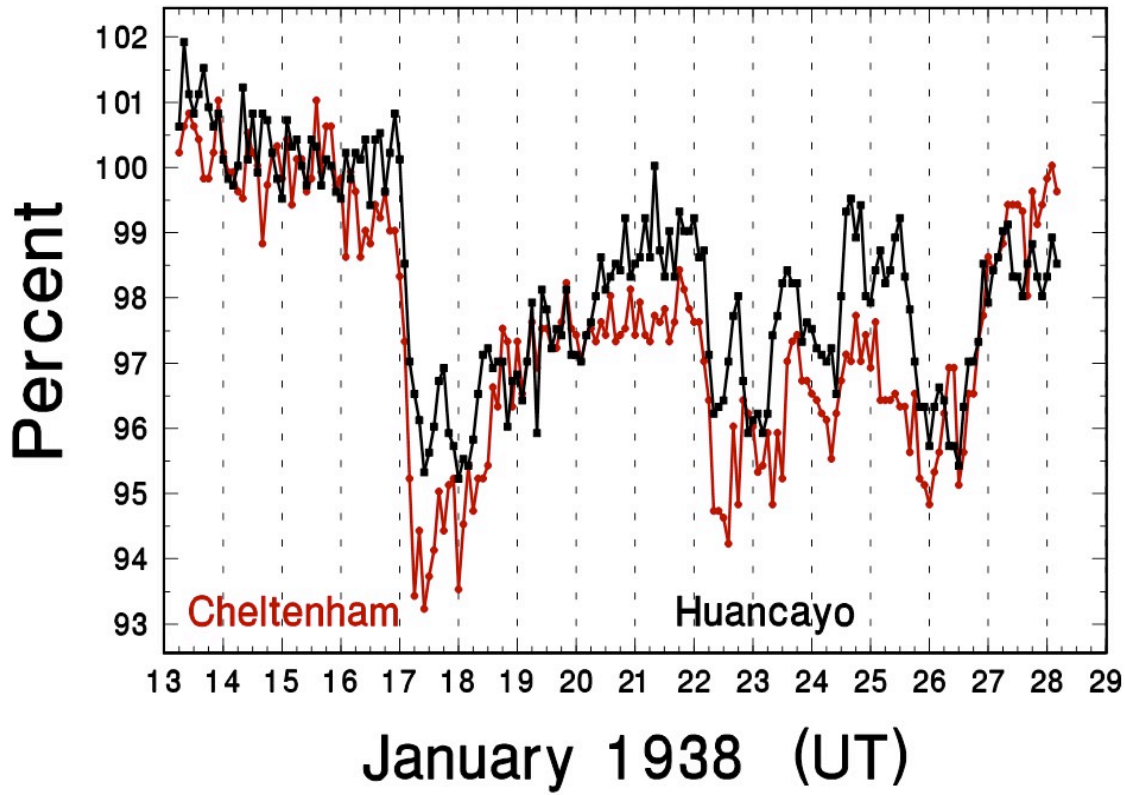

Figure 7: Bi-hourly variability of the cosmic-ray intensities recorded in the ionization chambers at Cheltenham and Huancayo during 13 – 28 January 1938, derived from Lange and Forbush (1948). Both LT and UT times are given as the photographic records, the Tables in Lange and Forbush (1948) and some of the figures in the publications of Forbush are in LT. Note that the cosmic ray intensity shown in Figure 2 of Forbush (1938) is an average of the intensities measured at Boston (*i.e.*, Cambridge), Cheltenham and Huancayo.

The published traces of cosmic-ray data (Forbush, 1938; Hess *et al.*, 1938) show Forbush Decreases during the geomagnetic storms on 17/18, 21/22, and 25/26 January. Figure 7 presents the cosmic ray intensity as recorded by the Cheltenham and Huancayo ionization chambers during 13 – 28 January. The largest decrease was on 17/18 January with more moderate decreases later in the month. This implies that the magnetic structure(s) responsible for the first geomagnetic storm provided a stronger barrier to cosmic rays than the later magnetic clouds. The double dip in the Dcx index (Fig. 6) associated with the first storm on 16 and 17 January is consistent with a strong shock passage, followed by an ICME with trailing southward field. A modern, but slower, analog for this event can be found in the shock/ICME passage of 12 – 14 October 2000 (see Fig. 3 of Richardson and Cane 2010) and associated indices in the OMNIWeb database. Furthermore, we suggest that the orientation of the ICME upon arrival at the Earth's magnetosphere was such that the southward component of the interplanetary magnetic field had a moderate magnitude, thus giving rise to a more moderate storm than the following ICMEs.

The geomagnetic storm on 17/18 January had two possible solar origins: From either the solar activity 04:40 – 05:30 UT on 14 January (Hα patrol) or 00:40 UT on 16 January with a solar flare effect (SFE). The latter event with a fast ICME (average velocity ≈ 1900 km/s) seems the most likely progenitor of the SSC at Earth at 22:36 UT and the resulting Forbush Decrease. This is because the large active region MWO 5726 was located around N18W30 at the time of the SFE (essentially the X-ray flare; see Figures 1 and 2) based on the Greenwich Photoheliographic Records and was more favorably located for a fast CME to travel to Earth producing the largest Forbush Decrease of this series of events.

The possibility of a solar proton event during this active period has been considered (*e.g.*, Bezprozvannaja, 1962). Interestingly, almost simultaneously, a complete blackout of short-wave communications was reported in the polar region during 16 – 19 January 1938 (Moisejev, 1939b). Švestka (1966) citing Bezprozvannaja (1962) includes 16 January 1938 among his list of polar blackout events, but Švestka did not include this event in his list of possible proton events. Besprozvannaya (1962) listed indirect data for an abnormal polar cap absorption on 16 January, but no time was given. Besprozvannaya's data were from vertical incidence ionospheric soundings from Tikhaya Bay (N80°19′, E052°47′). At 00:40 UT on 16 January 1938, the time of the SFE, Tikhaya Bay was under polar night with a maximum solar altitude of ≈ −25.5°. A SFE is indicative of a strong X-ray event, and while there is no solar activity listed in the QBSA (D'Azambuja, 1938), it is possible there were no solar optical observations at that time. A large solar X-ray event at 00:40 UT would not have impacted the ionosphere above Tikhaya Bay; however, a solar proton event with particles >10 MeV could easily produce a polar cap blackout in the dark polar regions as determined by simultaneous measurements of solar proton events and polar cap riometer absorption (Sellers *et al*., 1977; see also Shea and Smart, 2012).

Scanned copies of the original real-time photographic records from the Carnegie ionization chambers are available at the National Centers for Environmental Information, Data Services Division, Asheville, North Carolina in the USA. The Carnegie ionization chambers (IC) have been described previously (Compton *et al*., 1934). A brief summary is appropriate here to better understand a unique characteristic of these detectors: their high temporal resolution. The IC was an analogue instrument with an intrinsic time constant < 1 minute, and the time integral of the ionization current was recorded on a moving photographic strip with a temporal resolution of better than one minute[11]. Consequently, these detectors provided a detailed minute-by-minute record of the intensity of the cosmic radiation. Normally the trace rose or fell at a smooth essentially monotonic rate within the hour, except when a ground level enhancement (GLE) was detected immediately displaying an abrupt and subsequently strongly variable change in the slope of the trace. One of the authors of this paper (KGMcC) built and operated an IC and has examined at least 30000 hourly traces over his career including all the known GLEs between 1940-1960 as recorded by the Carnegie ICs. In all those 30000 records, there were no sharply variable traces other than the known GLEs in 1942, 1946, 1949, 1956 and 1960.

We have scanned the IC records from both the Cheltenham and Christchurch detectors for January 1938 and several subsequent months. During January 1938, both ionization chambers exhibited the relatively slow hour-by hour variations including the Forbush decreases shown in Figure 7. The standard deviation of an hourly increase in ionization current was estimated to be 1.0% in Tatel's Carnegie Institute Workbook. At 19:35 ± 2 UT the Cheltenham ionization current increased sharply within one minute varying thereafter as shown in Figure 8. Following a broad 30 minute maximum, it had disappeared by 22:00 UT. Close inspection of the photographic records shows a double pulse: the onset of the first increase at ≈ 19:35 UT (14:35 LT) and the second one at ≈ 20:25 UT (15:25 LT). Integrated over the period 19:35 – 22:00 UT the enhancement represents an increase in cosmic ray intensity of less than one standard deviation and could not be classed as a GLE on that basis. However our interpreters' experience with this now rare type of recording; the complete absence of any other similar sharp changes in 30,000 hourly records other than in GLE; and its strong similarity to the sharp temporal variations during known GLE leads us to record the possibility that it may be such in the analysis of this unusual series of solar events. This possibility is strengthened by the ionospheric and radio observations outlined below. We note, however that the Christchurch IC trace looked normal with little recognizable deviations from the norm. Therefore, there was no evidence of a correlating enhancement at Christchurch in the Southern Hemisphere at a similar geomagnetic cut-off to Cheltenham in the Northern Hemisphere, indicative of an anisotropy as is common in GLE.

---

[11] See Figure 6.2-1 of Beer *et al.*, (2012) for an example of a GLE as recorded by the Christchurch ionization chamber on 7 March 1942.

Hayakawa et al. (2020) The extreme storms in January 1938, *The Astrophysical Journal*, DOI: 10.3847/1538-4357/abc427

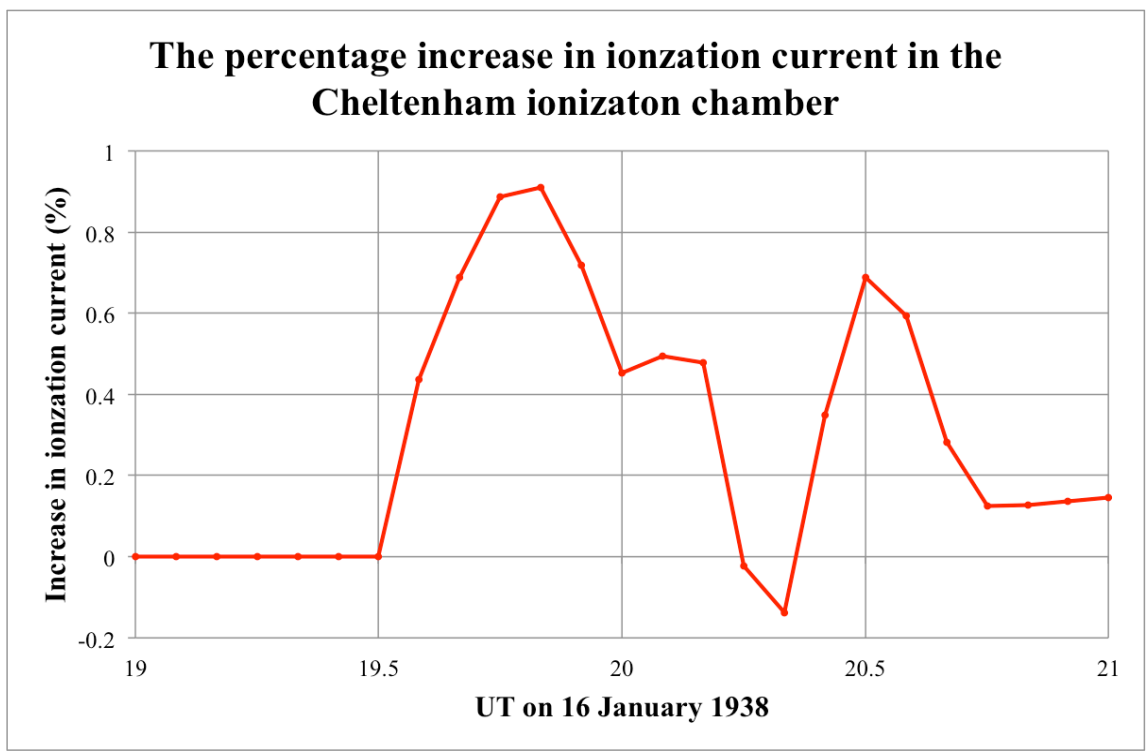

Figure 8: The percentage increase in ionization current in the Cheltenham ionization chamber, derived from the digital copies of the original real-time photographic records from the Carnegie ionization chambers preserved at the National Centers for Environmental Information (NCEI).

A small enhancement of the type shown in Figure 8 would not have been considered significant by Dr. Forbush whose main interest at that time was in geomagnetic field variations and whose supervisor was adamant that the sun could not accelerate cosmic rays to relativistic energies (S. Forbush's personal communication with KGMcC in 1962). The concept of a solar cosmic ray event did not occur until the large events in 1942, but it was not until 1946 when these *data and* their interpretation were published (Forbush, 1946).

An example of a small muon increase associated with a significant ground-level event was recorded during the complex array of multiple CMEs and interplanetary shocks in November 1960. For the GLE on 12 November 1960, the high sensitivity MIT muon detector only recorded a 1% increase compared to ≈ 60% for several neutron monitors (Steljes *et al*., 1961). That GLE and the following one (15 November 1960) were not discernible in the Cheltenham (Fredrichsburg) IC record. That is, on the basis of the evidence from muon detectors, the putative event on 16 January 1938 (Figure 8) was possibly more intense than the well-documented, substantial, and initial anisotropic GLE on 12

November 1960. While further evidence is needed to prove the small increase at Cheltenham between ≈ 19:35 and 20:45 UT on 16 January 1938 as a GLE, we note the following:

(1) Radio fade-outs occurred in the eastern USA from 16:40-20:20 UT on 16 January (Gilliland *et al.*, 1938)

(2) Approximately 18 hours after the Cheltenham increase, the disturbances in the geomagnetic field greatly intensified indicative of the possible arrival of yet another interplanetary shock during this highly disturbed period. There was also an additional short decrease in the galactic cosmic ray intensity that had just started to recover from the previous geomagnetic storm (see Figures 3 and 7).

This sequence seems more than just a coincidence. We can infer from all of the flare activity during 14 – 16 January that there were likely multiple ICMEs, possibly interacting and merging, on their way to Earth. The reported radio fadeouts between Ohio and Washington, DC from 16:40 to 20:20 (Gilliland, 1938) imply their cause as a long duration X-ray event or an influx of GeV protons, or their combination. This strongly suggests significant solar activity between ≈ 16:40 – 20:20 on 16 January. In order to reach the path between Ohio and Washington (≈ 50° – 55° MLAT, see Figure 4), the protons would have a rigidity in excess of 2 GV thus rending credit to a hypothesized high-energy particle event.

To summarize the sequence of activity, we know that an ICME was advancing close to Earth as evidenced by a SSC at 22:36 UT on 16 January. If another flare occurred around 19:35 UT with the acceleration of high-energy (GeV) protons, those protons would pass through the intervening shock front to arrive at Earth. The slightly less energetic particles, moving slower, would most likely be accelerated by the magnetic fields in the shock front of both the secondary ICME and the primary ICME. Such particles might have been trapped/scattered between the preceding merged interaction region and the new advancing CME as in the August 1972 sequence (*e.g.*, Knipp *et al.*, 2018).

We thus assess the enhancement at 19:35 UT on 16 January 1938 as a plausible — but not yet confirmed — GLE before the earliest known GLE reported on 28 February 1942 (Forbush, 1946), on the basis of consistent overall enhancement in the scaled data at Cheltenham. This finding may let us chronologically bridge the known GLEs (McCracken, 2007; Shea and Smart, 2019; Usoskin *et al.*, 2020b) and historical GLEs in 774/775 and 993/994 confirmed in the tree rings and ice cores (Miyake *et al.*, 2012, 2013; Usoskin *et al.*, 2013; Mekhaldi *et al.*, 2015), which is considered a factor of 50-100 stronger than the strongest known GLE of 23-Feb-1956 (*e.g.*, Mekhaldi *et al.*, 2015; Miyake *et al.*, 2019; Usoskin *et al.*, 2020a; Cliver *et al.*, 2020). Moreover, this extended time-scale is typical of a GLE originating from solar activity near and east of solar central meridian (McCracken and Palmeira, 1960; Stoker, 1995, p. 362). The GLE on 12 October 1981, associated with solar activity at E31, had an initial onset at 06:45 UT recorded by the Goose Bay neutron monitor in Canada and a broad maximum intensity of 8.5 % between 08:45 and 10:00 UT as shown in the GLE database at Oulu University[12] (Poluianov *et al.*, 2017; Gil *et al.*, 2018; Usoskin *et al.*, 2020b; D. F. Smart, private communication in 2020). This interpretation is consistent with the location of the flare-productive AR (MWO 5726) in the eastern solar disk at that time (Figure 1). Thus there are two possible source flares for a possible solar proton event on 16 January 1938: the SFE at 00:40 UT or unreported solar activity around 19:35 UT.

## 6. Discussions

Our study confirms the arrivals to Earth of three ICMEs in late January 1938, as implied by the occurrence of intense geomagnetic storms and Forbush Decreases (summarised in Table 2). The Sun was extremely active in this interval, probably because of hyper-productive ARs such as AR 5726 (RGO 12673). With a significant level of sunspot and flare activity during 14−22 January, the source region of the CME associated with the geomagnetic storm on 17/18 January is somewhat uncertain, hosting the two major Hα flares at 04:40 – 05:30 on January 14 (from AR 5726 at N10E45 and AR 5719 at N25W35) and two large SFEs at 17:07 on January 15 and at 00:40 on January 16 as its possible sources. The largest Forbush Decrease (≈ 6%) as recorded by ionization chambers on 17/18 January in this sequence indicates its source to be from a fast ICME originating from somewhere near the disk center. Therefore, we consider the SFE at 00:40 on 16 January as its probable source.

Table 2: Time series for the major solar flares, SSCs, Forbush Decreases, and geomagnetic storms in the interval of 14 – 26 January 1938, examined in this article. Here, we have used abbreviations of SFE (solar flare effect), PCA (polar cap absorption), GLE (ground level enhancement), SSC (storm sudden commencement), GMS (geomagnetic storm), FD (Forbush Decrease), RF (radio fadeout), and Kakioka ED (Kakioka Event Database).

---

[12] https//gle.oulu.fi

| Date | Time (UT) | Event | Magnitude | Reference | Notes |
|---|---|---|---|---|---|
| 14 | 0440-0530 | Flare | 3 | D'Azambuja 1938 | AR 5726 at N10E45 |
| 14 | 0440-0530 | Flare | 2+ | D'Azambuja 1938 | AR 5719 at N25W35 |
| 14 | 1629-1835 | RF | | Gilliland et al. 1938 | in the Eastern USA |
| 14 | 1840-1857 | RF | | Gilliland et al. 1938 | in the Eastern USA |
| 15 | 1707 | SFE | | Bartels 1939 | at Huancayo *etc*. |
| 15 | 1708-1724 | RF | | Gilliland et al. 1938 | in the Eastern USA |
| 15 | 1719-1737 | Flare | 2 | D'Azambuja 1938 | AR 5719 at N21W53 |
| 15 | 1838-1935 | RF | | Gilliland et al. 1938 | in the Eastern USA |
| 16 | 0040 | SFE | | Bartels 1939 | at Watheroo |
| 16 | 0040 | SFE | | Yokouchi 1953 | at Kakioka |
| 16 | 0044 | SFE | | Wadsworth 1938 | at Apia |
| 16 | | PCA | | Besprozvannaya 1962 | Indirect data |
| 16 | 1640-2020 | RF | | Gilliland et al. 1938 | in the Eastern USA |
| 16 | ~1900 | RF | | Spokane Daily Chr., 1938-01-17 | in the Western USA |
| 16 | 1935-2025 | GLE? | | Cheltenham IC | a plausible GLE |
| 16 | 2236 | SSC | 72 nT | Kakioka ED | |
| 17 | ~16 | GMS | −171 nT | Dcx index | |
| 17 | | FD | ~6% | Forbush 1938 | |
| 19 | 1115-1200 | Flare | 2 | D'Azambuja 1938 | AR 5726 at N15W25 |
| 19 | 2238-2312 | RF | | Gilliland et al. 1938 | in the Eastern USA |
| 20 | 1608-1800 | Flare | 2 | D'Azambuja 1938 | AR 5726 at N24W23 |
| 20 | 1820-2127 | Flare | 3 | D'Azambuja 1938 | AR 5726 at N18W30 |
| 20 | 1800-1840 | RF | | Gilliland et al. 1938 | in the Eastern USA |
| 20 | 1902-2040 | RF | | Gilliland et al. 1938 | in the Eastern USA |
| 21 | 1640-1920 | RF | | Gilliland et al. 1938 | in the Eastern USA |
| 22 | 0242 | SSC | 63 nT | Kakioka ED | |
| 22 | | GMS | −328 nT | Dcx index | |
| 22 | | FD | ~3% | Forbush 1938 | |
| 24 | 0100-0130 | Flare | 2 | D'Azambuja 1938 | AR 5726 at N22W85 |
| 24 | 0300-0340 | Flare | 3 | D'Azambuja 1938 | AR 5726 at N22W85 |
| 24 | 0512-0700 | Flare | 3 | D'Azambuja 1938 | AR 5726 at N22W80 |
| 24 | 1810-1850 | RF | | Gilliland et al. 1938 | in the Eastern USA |
| 25 | 1151 | SSC | 63 nT | Kakioka ED | |
| 25 | ~23 | GMS | −336 nT | Dcx index | |

| | | | | |
|---|---|---|---|---|
| 25 | | FD | ~2% | Forbush 1938 | |

If this was the case, the initial ICME was the fastest with its average velocity of ≈ 1900 km/s, in contrast with the second and third ICMEs (≈ 1370 km/s and ≈ 1260 km/s, respectively). The amplitudes of the corresponding SSCs at Kakioka Observatory were unusual in terms of occurrence frequency on the basis of its regular observations in 1924 – 2013 (Figure 4 of Araki, 2014). As the SSC amplitude ($\Delta H$) is empirically described as $\Delta H = C\Delta P^{1/2}$, where $C$ = 15 nT/nPa$^{1/2}$ and $\Delta P$ means change in the solar wind dynamic pressure (Araki, 2014; see also Siscoe *et al*., 1968), the $\Delta P$ for each of these ICMEs is computed as ≈ 23 nPa, ≈ 18 nPa, and ≈ 18 nPa, respectively. The estimated dynamic pressures are much higher than those for the more common ICMEs and are therefore classified as extreme cases (*e.g*., Lugaz *et al*., 2015). The lower limit of the solar wind density can be estimated by the equation $n = \Delta P/mV^2$, where $m$ means the solar wind mass (≈ 1.16 times proton mass), $n$ means the solar wind density, and $V$ means solar wind velocity (Araki, 2014). On this basis, the lower limit of the solar wind density of these ICMEs is roughly estimated ≈ 3 cm$^{-3}$, ≈ 5 cm$^{-3}$, and ≈ 6 cm$^{-3}$, respectively. Additionally, the dynamic pressures of the solar wind before the shock arrivals are unknown and they probably make the actual solar wind density slightly higher than our estimate.

For the first storm, the SSC amplitude is greatest, and the magnitude of Forbush Decrease is maximum (Table 1). We consider that the initial ICME was the strongest in terms of speed and dynamic pressure of the solar wind as well as the magnitude of IMF in comparison with the second and third ICMEs. Nevertheless, the magnitude of the first geomagnetic storm was weaker (Dcx ≈ −171 nT) in comparison with the other two (≈ −328 nT and ≈ −336 nT), unlike the magnitudes of the Forbush Decreases for these ICMEs (≈ 6% vs ≈ 3% and ≈ 2%). This contrast can be explained by the weaker southward component of the IMF in the initial ICME. It should be also noted that their variable impact angles might have played another role. The solar wind density sometimes contributes to the intensification of the ring current, in addition to the solar wind speed and the southward component of the IMF (Thomsen *et al*., 1998; Smith *et al*., 1999; O'Brien and McPherron, 2000; Ebihara and Ejiri, 2000). However, the intensity of the ring current is suggested to be not proportional to the solar wind density because of the shielding electric field that impedes the intensification of the ring current (Ebihara *et al*., 2005).

On the basis of the sunspot positions (N18W30, N22W85, and N22W80), it is also speculated that only part of the ICMEs impacted the terrestrial magnetosphere upon the first and third SSCs (see Cliver, 2006; Gopalswamy *et al*., 2005, 2007). The first ICME had at least three scenarios and the ICMEs of initial two eruptions (at 04:40 – 05:30 on 14 January and 17:07 on 15 January) may have cleared the path for the ICME for the ICME of the third eruption (at 00:40 on 16 January). Similar cases are found in the Hydroquebec storm in March 1989 and the Halloween sequence in 2003 October. Around the Hydroquebec storm on 13/14 March 1989, AR 5395, a large and complex δ-type sunspot group, caused a sequence of flares including 11 X-class flares such as the March 6 flare (X15) and March 17 flare (X6.5) (Allen *et al*., 1989; Boteler, 2019). Similarly, AR 10486 caused a series of flares and launched a series of ICMEs, which resulted in major geomagnetic storms of the Halloween storm category (Gopalswamy *et al*., 2005, 2007; Lefevre *et al*., 2016; Shiota and Kataoka, 2016).

This stormy sequence caused a series of space weather hazards in January 1938 (*e.g*., Lennaham, 1938). As described above, the USSR polar region witnessed a complete blackout of short-wave communications on 16 – 19 January 1938. In the American sector, the *New York Herald Tribune* (1938-01-26, p. 21) and the *Phoenix Arizona Republic* (1938-01-26, p. 1) reported radio disruptions even troubling airline flight operations on Monday (January 17), Friday (January 21), and the date immediately before the article release (25 January). During and after the sudden storm commencement on 16 January, Western Union Telegraph Company and American Telephone and Telegraph each reported strong effects of geomagnetically induced currents in the eastern portion of the US (The Christian Senior Monitor, 1938). Western Union Telegraph Company reported slight effects of geomagnetically induced currents beginning on 22 January as well. The problems increased in intensity until the afternoon of 25 January, when maximum readings exceeding 400 volts were recorded in many parts of the United States (Willever, 1938). Dr. Dinsmore at Griffith Observatory reportedly saw 'one large gas eruption or prominence' in connection with the great sunspot on Monday (January 24). Dr. Frederick Seares of the Mt. Wilson Observatory noted, "The radio fade-outs are caused by energy which comes from one of the gas eruptions to Earth with the speed of light." This could be associated with the limb eruption of the ICME around the fourth flare on the same date (24 January). With caveats on the chronological uncertainty, the flare reported in Australia, the erupting prominence, the geomagnetic storms, and the radio blackouts affecting airlines are all connected. Radio disruptions can be from flares, radio bursts or CME-driven geomagnetic/ionospheric storms as shown for the May 1967 storm (Knipp *et al*., 2016). As such,

Hayakawa et al. (2020) The extreme storms in January 1938,
*The Astrophysical Journal*, DOI: 10.3847/1538-4357/abc427these reports in the Tribune and Arizona Republic show early parallels to the May 1967 storm, appearing in the January 1938 literature, however apparently less dramatically.

## 7. Conclusions

In this article, we have evaluated the time series of the solar-terrestrial storms in January 1938. These major storms occurred around the maximum of Solar Cycle 17, quite consistently with the existing statistics of occurrence frequency of such space weather events (Lefevre *et al.*, 2016). Here, we have located four major flares with importance of "3" in Hα flaring area on 14 (4:40 – 5:30 UT), 20 (18:20 – 21:27 UT), and 22 (3:00 – 3:40 UT and 5:12 – 7:00 UT) January, from a giant AR of RGO 12673 (≤ 3627 msh). We also identified a large SFE at 00:40 UT on 16 January, which was presumably launched from the same AR.

These flares were followed by three major storms and therefore indicate at least three major ICMEs in this interval. Three SSCs show their arrival time on 16 (22:36 UT, 72 nT), 22 (02:42 UT, 63 nT), and 25 (11:51 UT, 63 nT) January, as observed at Kakioka Observatory. Taking the time of flare as the CME eruption and the time of the SSCs as the time of arrival at Earth, we have evaluated the parameters of these ICMEs: their average velocity as ≈ 1900 km/s (first ICME), 1370 km/s (second ICME), and 1260 km/s (third ICME); their solar wind dynamic pressure as ≈ 23 nPa, ≈ 18 nPa, and ≈ 18 nPa; and lower limit of their solar wind density as ≈ 3 $cm^{-3}$, ≈ 5 $cm^{-3}$, and ≈ 6 $cm^{-3}$, respectively. Our estimates classify these ICMEs into the extreme category (*e.g.*, Lugaz *et al.*, 2015).

The intensity and evolution of these three major storms have been evaluated with the Dcx index, which was calculated on the basis of mid-latitude magnetograms. The storm intensities and their peaks have been estimated as Dcx ≈ −171 nT at 16 UT on January 17, Dcx ≈ −328 nT at 11 UT on 22 January, and Dcx ≈ −336 nT at 23 UT on 25 January. Accordingly, it is shown that two major storms, which are almost comparable with each other, followed a less intense storm.

During the storms on 21/22 and 25/26 January, great auroral displays were globally reported. Investigating aurorae visible at low-latitudes (|λ| < 40°), we have located the equatorial boundary of the auroral visibility down to 29.5° MLAT and 29.9° MLAT during these two storms on 21/22 and 25/26 January. On the basis of the records with elevation angle, we have also reconstructed the equatorial boundaries of the auroral oval for these storms as ≈ 40° ILAT for each. This also shows that these two storms rivaled with each other unlike what has been discussed so far (Silverman, 2006) and enhance the role of the 21/22 January more than has been previously considered.

Comparing the time series of the magnetic disturbance and the visibility of the low-latitude aurorae, we have revealed that the low-latitude aurorae ($|\lambda| < 40°$) were reported around the peak of the geomagnetic storms (Dcx ≤ −200 nT). The aurorae on 21/22 January were most seen around Japan as it peaked at 11 UT (late evening in the Japanese sector). On the other hand, the aurorae on 25/26 January were most seen around the Mediterranean Sea, as it peaked at 23 UT (close to midnight in the European sector).

In combination with the first less-intense storm on 17 January, it is suggested that the first massive and fast ICME swept the interplanetary space, and allowing the following two more major ICMEs to be much more geo-effective despite their relatively more moderate dynamic pressure. These cases show that a sequence of ICMEs makes its consequences much more serious to the modern technological infrastructure than a single major ICME or a resultant geomagnetic storm. Our results show that a "perfect storm" was formed in mid January 1938, just like a sequence of the major geomagnetic storms associated with the so-called "Halloween storm" in October 2003 and other perfect storms in July 2017, July 2012, and August 1972, where large sunspot active regions are capable of launching multiple ICMEs in rapid sequences and create extreme geomagnetic storms as a collective action of several lesser in intensity storms (Cliver and Svalgaard, 2004; Gopalswamy *et al*., 2005; Knipp *et al*., 2018; Liu *et al*., 2019).

The contemporary ionization chambers measurements of the cosmic ray intensity have recorded three major Forbush Decreases upon the arrival of these three ICMEs. These data show that the initial decrease was the largest (≈ 6%) in accordance with their SSC amplitudes and in contrast with their Dcx intensities. This indicates the initial geomagnetic storm was likely caused by a faster ICME than the following ICMEs and makes the SFE at 00:40 UT on January 16 as its probable source. Contemporary records of the polar cap absorption indicate the occurrence of a significant solar proton event on January 16. This could be associated with the SFE at 00:40 on this date. Otherwise, close inspection of the original traces of the Cheltenham ionization chamber measurements reveals a small double pulse at ≈ 19:35 UT and ≈ 20:25 UT indicating the potential occurrence of a weak GLE on this date, thus predating the earliest known GLE in the observational history. This plausible – but not confirmed – GLE may also be the source for the polar cap absorption.

**Acknowledgments**

This work was supported in part by Grant No. 19-02-00088 by RFBR, projects HISTIGUC (PTDC\FER-HFC\30666\2017), MAG-GIC (PTDC/CTA-GEO/31744/2017), JSPS Grant-in-Aids JP15H05812, JP17J06954, JP20K22367, JP20K20918, and JP20H05643, JSPS Overseas Challenge Program for Young Researchers, the 2020 YLC collaborating research fund, and the research grants for Mission Research on Sustainable Humanosphere from Research Institute for Sustainable Humanosphere (RISH) of Kyoto University and Young Leader Cultivation (YLC) program of Nagoya University. We thank Afroditi Nasi, Varvara Kotsiourou, and Kosuke Fukuda for helping us interpreting the contemporary Greek newspapers; and Chiaki Kuroyanagi for helping us interpreting the contemporary USSR and Eastern European newspapers. We thank Don F. Smart for important advice and discussions on interpretations of the ionization-chamber data. We thank Murray Parkinson, Helen Fischeler, John Kennewell, Michael Wheatland, and Donald Melrose for their helpful advice on the Australian flare patrol records. We thank Roger Ulrich for his helpful advice on the MWO flare patrol records. We thank Ian Richardson for his helpful discussions on Dst profile for shock of ICMEs. HH thanks Ilya G. Usoskin for his helpful comments on the early cosmic-ray reports, Sam M. Silverman for his helpful advice on the auroral sightings in the West Europe and North America, José R. Ribeiro and Ana Correia for their helpful discussions on the Iberian auroral sightings, Aki Machida and Koji Mikami for their advice on the archival materials in the Japan Meteorological Agency, and Denny M. Oliveira and Sean P. Blake for their helpful discussions on the ICMEs. HH also thanks Solar Science Observatory of the NAOJ for providing copies of *Quarterly Bulletin on Solar Activity* and Shingo Nagamachi for his advice on the observational data at Kakioka Observatory. HH has been benefited from discussions within the ISSI International Team #510 (SEESUP Solar Extreme Events: Setting Up a Paradigm) and ISWAT-COSPAR S1-02 team. AAP thanks the GOOGLE Books Team for promptly reviewing the request and releasing the electronic version of the Bulletin of the Astronomical-Geodetical Society of the USSR, Issues 1-4. AAP is a member of the international team of Modeling Space Weather and Total Solar Irradiance over the Past Century supported by the International Space Science Institute (ISSI), Bern, Switzerland and ISSI-Beijing, PRC. DJK was partially supported by AFOSR grant No: FA9550-17-1-0258. CITEUC is funded by Portuguese Funds through FCT (project: UID/MULTI/00611/2019)

**Data Availability**

The SSCs and geomagnetic storms recorded at Kakioka Observatory have been acquired from the Kakioka Event Database. The results presented in this paper (Figure 3) use Dcx indices provided by the Dcx server of the University of Oulu, Finland (http://dcx.oulu.fi). The auroral data in this article has been summarized in https://www.kwasan.kyoto-u.ac.jp/~hayakawa/data/1938. The hourly magnetic measurements and the standard Dst index have been acquired from the WDC for geomagnetism at Kyoto. The newspapers in Australia and New Zealand are consulted in the collections of the National Libraries of Australia and New Zealand. We thank Shawn Hardy of the Carnegie Institute for carefully preserving the original Forbush ionization chamber records, the National Geophysical Data Center in Boulder, Colorado for digitizing many of the original records, and the National Centers for Environmental Information, Data Services Division, Ashville, North Carolina for archiving these original records. The modern GLE data captured in the neutron monitors have been acquired from the GLE Database of the University of Oulu (https://gle.oulu.fi).